\documentclass{elsarticle}
\usepackage[usenames,dvipsnames]{color}
\usepackage{graphicx}
\usepackage{amsmath}

\begin{document}
\begin{frontmatter}
\title{Negative mobility of a Brownian particle: \\ strong damping regime}
\author{A. S{\l}apik}
\author{J. {\L}uczka\corref{1}}
\ead{jerzy.luczka@us.edu.pl}
\author{J. Spiechowicz}
\address{Institute of Physics, University of Silesia, 40-007 Katowice, Poland}
\address{Silesian Center for Education and Interdisciplinary Research, University of Silesia, \\41-500 Chorz{\'o}w, Poland}
\begin{abstract}
We study impact of inertia on directed transport of a Brownian particle under non-equilibrium conditions: the particle moves in a one-dimensional periodic and \emph{symmetric} potential,  is driven by both an unbiased time-periodic force and a constant force, and is coupled to a thermostat of temperature $T$. Within selected parameter regimes this system exhibits  negative mobility, which means that the particle moves in the direction opposite to the direction of the constant force. It is known that in such a setup the inertial term  is \emph{essential} for the emergence of negative mobility and it cannot be detected in the limiting case of  overdamped dynamics. We analyse inertial effects and show that negative mobility can be observed even in the strong damping regime.    We determine the \emph{optimal} dimensionless mass for the presence of negative mobility  and reveal three mechanisms standing behind this anomaly: deterministic chaotic, thermal noise induced and deterministic non-chaotic. The last origin has never been reported. It may provide guidance to the possibility of observation of negative mobility for  strongly damped  dynamics which is of fundamental importance from the point of view of biological systems, all of which \emph{in situ} operate in fluctuating environments.
\end{abstract}
\end{frontmatter}
\section{Introduction}
When a system at thermal equilibrium is exposed to a weak external static force, its response is in the same direction as this of applied bias towards a new equilibrium. This restriction is no longer valid under nonequilibrium conditions when already an unperturbed system may exhibit a current due to the ratchet effect \cite{denisov2014}. Another example is the seemingly paradoxical situation of the negative mobility phenomenon when the system response is opposite to the applied constant force \cite{hanggi2009}. Such anomalous transport behaviour was predicted theoretically in 2007 in a system consisting of an {\it inertial} Brownian particle moving in a one-dimensional periodic symmetric potential \cite{machura2007}. Within a year of this  discovery,  negative mobility was confirmed experimentally in the experiment involving determination of current-voltage characteristics of the microwaved-driven Josephson junction \cite{nagel2008}. Yet further examples of this phenomenon has been described theoretically in companionship of coloured noise \cite{kostur2009}, white Poissonian noise \cite{spiechowicz2013jstatmech,spiechowicz2014pre}, dichotomous process \cite{spiechowicz2016jstatmech} and for Brownian motion with presence of time-delayed feedback \cite{hennig2009,du2011}, non-uniform space-dependent damping \cite{mulhern2013} and potential phase modulation \cite{dandogbessi2015}. Other illustrations include a vibrational motor \cite{du2012}, two coupled resistively shunted Josephson junctions \cite{januszewski2011,machura2012}, active Janus particles in a corrugated channel \cite{ghosh2014}, entropic electrokinetics \cite{malgaretti2014} as well as nonlinear response of inertial tracers in steady laminar flows \cite{sarracino2016}.

Modelling systems and understanding their generic properties discloses which components of the setup are crucial and which elements may be sub-relevant. For instance, transport in the micro-world is strongly influenced by fluctuations and random perturbations. In some systems, like biological cells  \cite{bressloff2013}, they can even play a dominant role and a typical situation is that motion  of particles  is strongly damped. This fact justifies the use of an overdamped dynamics for which the particle inertial term $M\ddot{x}$  can be formally neglected in comparison to the dissipation  term $\Gamma \dot x$ 
($M$ is the particle mass, $\Gamma$ is the friction coefficient and dot denotes a differentiation with respect to time $t$). Omission of the inertial term enormously simplifies the modelling and in many cases allows for an analytical solutions of the corresponding Fokker-Planck equation. However, properties and features which are allowed to occur in systems with inertia can completely disappear when the inertial term is put to zero. Certainly a more correct approach in such a situation is to include the inertial term and use a technique of mathematical sequences of smaller and smaller dimensionless  mass.  Our main objective is to investigate impact of  inertia on  negative mobility of a Brownian particle moving in  one-dimensional periodic systems. It is known that in such setups the inertial term is one of the key ingredients for the occurrence of this form of anomalous transport \cite{speer2007,spiechowicz2016pla} and negative mobility  is absent for the overdamped dynamics when $M\ddot{x}=0$ . We address the question whether it is still  possible to observe the negative mobility phenomenon in   strongly dissipative systems. In doing so, we first formulate the model and introduce the quantities of interest. Then we investigate the general transport behaviour as a function of model parameters and detect the optimal  dimensionless mass for the presence of negative mobility. In the next part we demonstrate three mechanisms responsible for the emergence of this anomalous transport phenomenon: deterministic chaotic, thermal noise induced and deterministic non-chaotic. Finally, we discuss impact of inertia on the directed long time particle velocity and provide some conclusions.

\section{Model}
The model of a Brownian particle moving in a one-dimensional periodic landscape has been already well established in the literature \cite{risken}. It has been used to explore a wide range of phenomena including ratchet effects \cite{spiechowicz2014prb,spiechowicz2015njp}, noise induced transport \cite{luczka1995, spiechowicz2015physscripta}, the negative mobility \cite{machura2007}, the enhancement of transport \cite{spiechowicz2014pre} and diffusion \cite{spiechowicz2015chaos}, the anomalous diffusion \cite{spiechowicz2015pre,spiechowicz2016scirep} and the non-monotonic temperature dependence of diffusion \cite{spiechowicz2016njp,spiechowicz2017chaos}. Here, we consider exactly the same model as in \cite{machura2007}:  a classical inertial Brownian particle of mass $M$, which  moves in a spatially periodic potential $U(x) = U(x + L)$ of period $L$ and is subjected to both an unbiased time-periodic force $A\cos{(\omega t)}$ of amplitude $A$ and angular frequency $\Omega$ and an external static force $F$. Dynamics of such a particle is described by the following Langevin equation \cite{machura2007}
\begin{equation}
\label{model}
	M\ddot{x} + \Gamma\dot{x} = -U'(x) + A\cos{(\Omega t)} + F + \sqrt{2\Gamma k_B T}\,\xi(t),
\end{equation}
where prime denotes a differentiation with respect to the particle coordinate $x$. Thermal fluctuations due to the coupling of the particle with the thermal bath of temperature $T$ are modelled  by Gaussian white noise of zero mean and unity intensity, namely
\begin{equation}
	\langle \xi(t) \rangle = 0, \quad \langle \xi(t)\xi(s) \rangle = \delta(t-s).
\end{equation}
The noise intensity factor $2\Gamma k_B T$ (where $k_B$ is the Boltzmann constant) follows from the fluctuation-dissipation theorem \cite{marconi2008} and ensures the canonical equilibrium Gibbs state when  \mbox{$A = 0$} and $F = 0$. The potential $U(x)$ is assumed to be in a \emph{symmetric} form with the period $L$ and the barrier height $2\Delta U$, namely, 
\begin{equation}
	U(x) = \Delta U \sin \left(\frac{2\pi}{L}x\right).
\end{equation}
There exists a wealth of physical systems that can be described by the Langevin equation (\ref{model}). An important cases that come to mind are the semiclassical dynamics of a phase difference across a resistively and capacitively shunted Josephson junction \cite{blackburn2016} and a cold atom moving in an optical lattice \cite{denisov2014, lutz2013}. Other examples include superionic conductors \cite{fulde1975}, dipoles rotating in external field \cite{coffey2004}, charge density waves \cite{gruner1981} and adatoms on a periodic surface \cite{guantes2001}.

\subsection{Scaling and dimensionless Langevin equation}

Since only relations between scales of length, time and energy are relevant for the observed phenomena, not their absolute values, we next formulate the above presented equation of motion in its dimensionless form. This can be achieved in several ways \cite{machura2008}. Because investigation of impact of the particle inertia on the system dynamics is our main goal, in the present consideration we propose the use of the following scales as the characteristic units of length and time \cite{machura2008}
\begin{equation}
	\hat{x} = \frac{x}{L}, \quad \hat{t} = \frac{t}{\tau_0}, \quad \tau_0 = \frac{\Gamma L^2}{\Delta U}.
\end{equation}
Under such a procedure the Langevin equation (\ref{model}) takes the dimensionless form \cite{spiechowicz2016njp}
\begin{equation}
\label{dimlessmodel}
	m\ddot{\hat{x}} + \dot{\hat{x}} = -\hat{U}'(\hat{x}) + a\cos{(\omega \hat{t})} + f + \sqrt{2D} \hat{\xi}(\hat{t}).
\end{equation}
In this scaling, the dimensionless mass is 
\begin{equation}
\label{mas}
m = \frac{\tau_1}{\tau_0} = \frac{M\Delta U}{\Gamma^2 L^2},  
\end{equation}
where the second characteristic time is $\tau_1 =M/\Gamma$ and the dimensionless friction coefficient is $\gamma =1$. Other parameters are: \mbox{$a = (L/\Delta U)A$}, $\omega = \tau_0\Omega$, \mbox{$f = (L/\Delta U)F$}. The rescaled thermal noise reads $\hat{\xi}(\hat{t}) = (L/\Delta U)\xi(t) = (L/\Delta U)\xi(\tau_0\hat{t})$ and assumes the same statistical properties as $\xi(t)$, namely $\langle \hat{\xi}(\hat{t}) \rangle = 0$ and $\langle \hat{\xi}(\hat{t})\hat{\xi}(\hat{s}) \rangle = \delta(\hat{t} - \hat{s})$. The dimensionless noise intensity $D = k_BT/\Delta U$ is the ratio of thermal energy and half of the activation energy the particle needs to overcome the nonrescaled potential barrier. The dimensionless potential $\hat{U}(\hat{x}) = \sin(2\pi \hat x)$ possesses the period $\hat L = 1$ and the  barrier height $\Delta \hat U = 2$. From now on, we will use only the dimensionless variables and therefore, in order to simplify the notation, we will omit the \emph{hat} notation in the above equation.
\subsection{Quantities of interest}
In the present study we are particularly interested in the impact of inertia  on properties of directed transport of particles in the stationary state. In the dimensionless formulation (\ref{dimlessmodel}) it can be realized by changing  the dimensionless mass (\ref{mas}). The case $m=0$ corresponds to overdamped dynamics and the setting $m\ll 1$ represents the strong damping regime, which means that $\tau_1\ll \tau_0$. The characteristic time $\tau_1$ is obtained from a particular form of Eq. (\ref{model}), i.e. $M\dot v +\Gamma v=0$ and has the interpretation of the relaxation time of the velocity of the free Brownian particle. The parameter $\tau_0$ is extracted from the equation $\Gamma \dot x = -U'(x)$ which can be viewed as the characteristic time to travel a distance from a maximum of the potential $U(x)$ to its minimum in the overdamped case (it is not exactly this time which is infinite in the considered case but $\tau_0$ scales it). It is remarkable that parameters of the potential $U(x)$ such as its barrier height $\Delta U$ and period $L$ are crucial for controlling the regimes of weak or strong damping. For instance, if $M$ and $\Gamma$ are fixed and the system is in a weak damping regime $m\gg 1$, the transition to the strong damping case $m\ll 1$ can be achieved  by lowering the barrier height and lengthening the period of $U(x)$. We have checked that for values $m \sim 0.1$ and smaller the system (\ref{dimlessmodel}) can be considered to be in the strong damping regime.  

Due to the presence of the external time-periodic driving $a\cos{(\omega t)}$, as well as the friction term $\dot{x}$,  the particle velocity $\dot{x}(t)$ approaches a unique non-equilibrium asymptotic long time state, in which it is characterized by a temporally periodic probability density. This latter function has the same period as the driving $\mathsf{T} = 2\pi/\omega$ \cite{jung1993}. Therefore, the first statistical moment of the instantaneous particle velocity $\langle \dot{x}(t) \rangle$ assumes for an asymptotic long time regime the form of a Fourier series over all possible harmonics \cite{jung1993}
\begin{equation}
	\lim_{t \to \infty} \langle \dot{x}(t) \rangle = \langle v \rangle + v_{\omega}(t) + v_{2\omega}(t) + ...
\end{equation}
where $\langle v \rangle$ is the directed (time independent) velocity, while $v_{n\omega}(t)$ denote time periodic higher harmonics  of vanishing time-average over the fundamental period $\mathsf{T} = 2\pi/\omega$. The observable of foremost interest in this study is  the directed transport component $\langle v \rangle$, which due to the mentioned particular decomposition can be obtained in the following way
\begin{equation}
\label{directedvelocity}
\langle v \rangle = \lim_{t \to \infty} \frac{\omega}{2\pi} \int_t^{t + 2\pi/\omega} ds \, \langle \dot{x}(s) \rangle,
\end{equation}
where $\langle \cdot \rangle$ indicates  averaging over all realizations of thermal noise  as well as over initial conditions for the position $x(0)$ and the velocity $\dot{x}(0)$. The latter is obligatory for the deterministic limit $D \propto T \to 0$  when dynamics may be non-ergodic and results can be affected by  specific choice of initial conditions \cite{spiechowicz2016scirep}.

Due to the multidimensionality of the parameter space of the considered model, as well as its nonlinearity, the force-velocity curve $\langle v \rangle = \langle v \rangle(f)$ is typically a nonlinear function of the applied bias $f$. From the symmetries of the underlying Langevin equation (\ref{dimlessmodel}) it follows that this observable is odd as a function of the external static force $f$, i.e. $\langle v \rangle(-f) = -\langle v \rangle(f)$ and in  consequence $\langle v \rangle(f = 0) \equiv 0$ \cite{denisov2014}. This is in clear contrast to the case of a ratchet mechanism, which exhibits the finite directed transport $\langle v \rangle \ne 0$ even at the vanishing static bias when $f=0$
 \cite{hanggi2009}. Since the observable of our interest is symmetric around $f = 0$, we limit our consideration to the positive bias $f > 0$. Then, for sufficiently small values of the external force $f$ the directed transport velocity $\langle v \rangle$ is usually its increasing function. Such regimes correspond to the normal, expected transport behaviour. However, in the parameter space there are also regimes for which the particle moves on average in the direction opposite to the applied bias, i.e. $\langle v \rangle < 0$ for $f > 0$, exhibiting anomalous transport behaviour in the form of the negative mobility phenomenon \cite{machura2007,speer2007}. 
 It has been already shown that there are two fundamentally various mechanisms responsible for  negative mobility in this setup, (i) generated by  chaotic dynamics and (ii) induced by thermal equilibrium fluctuations \cite{machura2007}. The latter situation is nevertheless rooted in the sophisticated evolution of the corresponding deterministic system described by Eq. (\ref{dimlessmodel}) with $D = 0$. Its  three-dimensional phase space $\{x, \dot{x}, \omega t\}$ is minimal for chaotic evolution, which is     important  for  negative mobility to occur. 

For the considered deterministic system with $D = 0$ there are three Lyapunov exponents $\lambda_1$, $\lambda_2$ and $\lambda_3$. It can be easily checked that the system is dissipative, i.e. the phase space volume is contracting during the time evolution. Therefore the sum of all Lyapunov exponents must be negative \cite{gollub}
\begin{equation}
	\lambda_1 + \lambda_2 + \lambda_3 < 0.
\end{equation}
One of the exponents, say $\lambda_3=0$ and the other, say $\lambda_2<0$. If the system is chaotic $\lambda_1$ must be  positive indicating divergence of the trajectories. Therefore to detect chaotic behaviour of the system it is sufficient to calculate the maximal Lyapunov exponent $\lambda=\lambda_1$ and check whether it is larger than zero \cite{boffetta2002}.

\section{Numerical simulation} 

Unluckily, the Fokker-Planck equation corresponding to the Langevin equation (\ref{dimlessmodel}) cannot be handled by any known analytical methods. For this reason, in order to analyse transport properties of the system, we carried out comprehensive numerical simulations. We integrated the Langevin equation (\ref{dimlessmodel}) by employing a weak version of the stochastic second order predictor corrector algorithm with a time step typically set to about $10^{-2} \times 2\pi/\omega$. We chose the initial coordinates $x(0)$ and velocities $\dot{x}(0)$ equally distributed over the intervals $[0,1]$ and $[-2,2]$, respectively. The quantities of interest were ensemble averaged over $10^3-10^4$ different trajectories, which evolved over $10^3-10^4$ periods of the external harmonic driving. All numerical calculations were performed by the use of CUDA environment implemented on a modern desktop GPU. This gave us possibility to speed up the computations up to a factor of the order $10^3$ times as compared to a common present day CPU method. Details on this promising scheme can be found in Ref. \cite{spiechowicz2015cpc}.
\begin{figure}[t]
	\centering
	\includegraphics[width=1.0\linewidth]{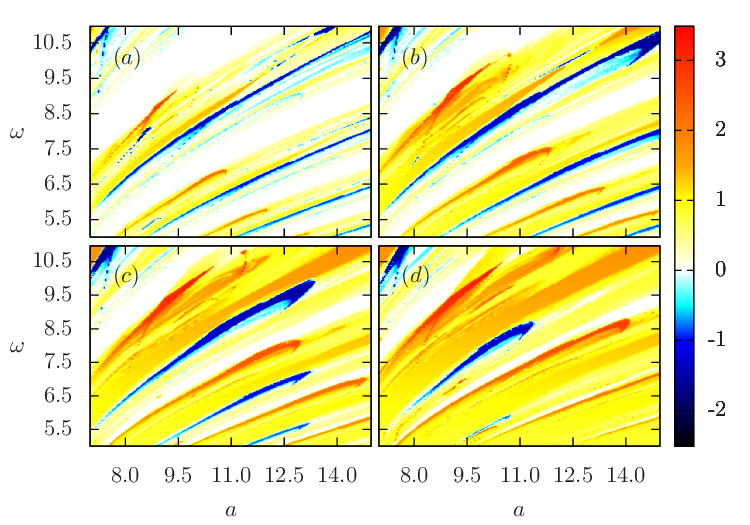}
	\caption{The Brownian particle asymptotic long time directed velocity $\langle v \rangle$ as a function of the amplitude $a$ and the angular frequency $\omega$ of the external unbiased harmonic driving $a\cos{(\omega t)}$ is shown for different values of the bias $f$ with $D = 0$ and $m = 0.1$. Panel (a) $f = 0.2$, (b) $f = 0.4$, (c) $f = 0.6$, (d) $f = 0.8$.}
	\label{fig1}
\end{figure}
\begin{figure}[t]
	\centering
	\includegraphics[width=1.0\linewidth]{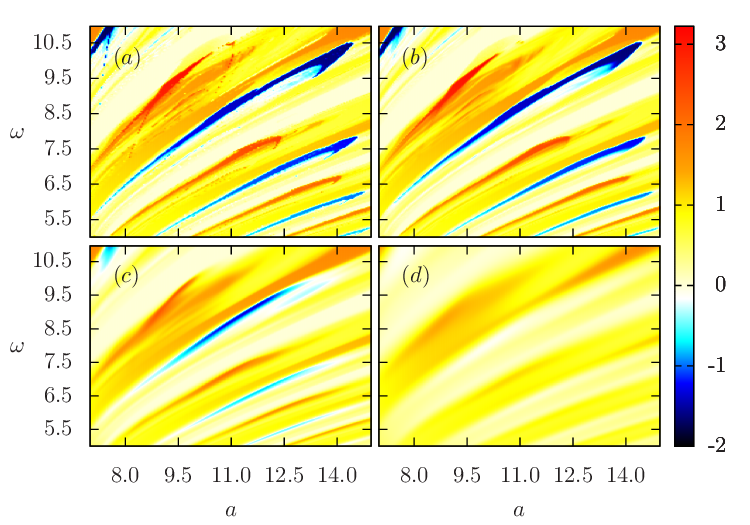}
	\caption{The directed velocity $\langle v \rangle$ versus the amplitude $a$ and the angular frequency $\omega$ is depicted for different values of thermal noise intensity $D$ with $m = 0.1$ and $f = 0.5$. Panel (a) $D = 0$, (b) $D = 10^{-5}$, (c) $D = 10^{-3}$, (d) $D = 10^{-2}$.}
	\label{fig2}
\end{figure}

Dynamics described by Eq. (\ref{dimlessmodel}) is characterized by a 5-dimensional parameter space $\{m, a, \omega, f, D\}$, the detailed exploration of which is a very challenging task even for our innovative computational method. However, we focus on the impact of the particle inertia on the anomalous transport processes occurring in this setup. This task is very tractable numerically with the currently available hardware. We start our analysis by looking at the deterministic system $D = 0$. We set the bias to a low value $f = 0.5$ and check how the directed velocity $\langle v \rangle$ depends on the remaining parameters. In doing so we performed scans of the following area $m \times a \times \omega \in [0.01,10] \times [0,20] \times [0,20]$ at a resolution of 200 points per dimension to determine the general behaviour of the system. Our results reveal that negative mobility is not present for $\omega > 18$ and $\omega < 2$. This is in agreement with the approximate solutions of Eq. (\ref{dimlessmodel}). In the limit of low frequencies an adiabatic approximation is valid \cite{chi1990}, while for high frequencies a solution can be formulated in terms of Bessel functions \cite{kautz1996}. Moreover, there is no net transport for $a < 4$.

\begin{figure}[t]
	\centering
	\includegraphics[width=1.0\linewidth]{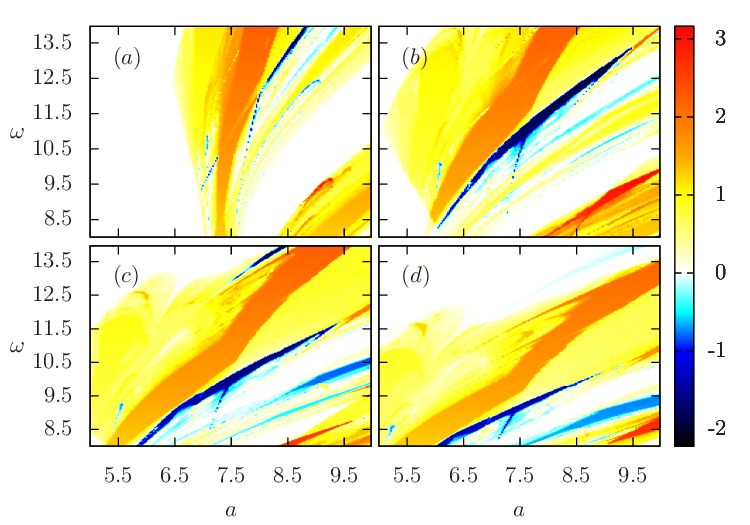}
	\caption{The directed velocity $\langle v \rangle$ versus the amplitude $a$ and the angular frequency $\omega$ is presented for different values of the particle mass $m$ with $D = 0$ and $f = 0.5$. Panel (a) $m = 0.05$, (b) $m = 0.1$, (c) $m = 0.15$, and (d) $m = 0.2$.}
	\label{fig3}
\end{figure}

\section{Results}
\subsection{General behaviour of the system}

The study of various aspects of transport in the system (\ref{dimlessmodel}) has been presented elsewhere  \cite{machura2007,speer2007}. Here, we focus our analysis primarily on the relationship between inertia and negative mobility.  
In Fig. \ref{fig1} we present the asymptotic long time directed velocity $\langle v \rangle$ depicted as a function of the amplitude $a$ and angular frequency $\omega$ of the external harmonic driving $a\cos{(\omega t)}$, for the strongly damped Brownian particle with the dimensionless mass $m = 0.1$ and different values of the external bias $f$. Surprisingly, despite the fact that the  inertia is one order  smaller than the dimensionless friction coefficient $\gamma = 1$, there are regions in the parameter space of the system where  negative mobility occurs. They form a band-like structure. The stripes of negative velocity are interspersed with the stripes of positive velocity and the difference in its magnitude in the neighbouring regions can be significant. This suggest that the system may be very sensitive to a small change of values of parameters. It can be observed that larger values of $f$ lead to reduction of the negative mobility areas towards the lower $\omega$ and $a$. At the same time the bands of negative velocity become wider and more intense. In contrast, the regions of positive mobility increase population and supersede their negative counterparts. Overall, as the bias $f$ is increased the band structure seems to zoom out until the directed velocity points to the direction given by $f$ and becomes almost constant in the whole map. In Fig. \ref{fig2} we depict the same characteristic but for different values of thermal noise intensity $D$ with fixed $m = 0.1$ and  $f = 0.5$.
One can expect that thermal  noise perturbs  deterministic dynamics. We observe that larger areas of negative mobility are relatively stable with respect to increasing temperature, while the smaller ones disappear more quickly. Thermal noise first blurs the band-like structure of negative mobility areas, erasing the finer details of the regions visible in the deterministic case $D = 0$. It seems to be obvious since  thermal noise enables random transitions between deterministically coexisting basins of attraction. For high enough temperatures,  negative mobility disappears completely. A careful inspection of Fig. \ref{fig2} reveals that there are regions in the parameter space where the directed velocity $\langle v \rangle$ is zero or positive in the deterministic case, but becomes negative upon the introduction of noise. This fact suggests that thermal fluctuations may induce  negative mobility or reverse its sign even for the strongly damped Brownian particle. Finally, in \mbox{Fig. \ref{fig3}} we present the directed velocity $\langle v \rangle$ versus the amplitude $a$ and the angular frequency $\omega$ for different values of the  mass $m$ with $D = 0$ and $f = 0.5$. The stripes of negative mobility move towards lower values of $\omega$ and higher values of $a$ as $m$ is increased. The band-like structure changes its inclination and the regions becomes more horizontally oriented. Moreover, for larger masses $m$ some new negative mobility bands appear, while at the same time the negative velocity tends to disappear in other regions. This effect suggest that there should exist an optimal mass $m$ for which the occurrence of negative mobility is mostly pronounced. 

\begin{figure}
	\centering
	\includegraphics[width=1.0\linewidth]{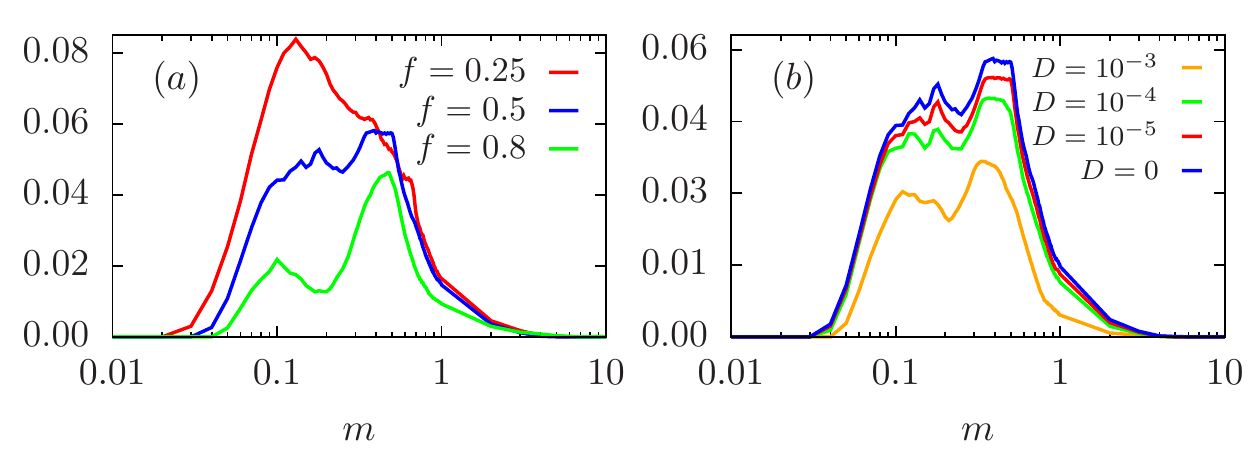}
	\caption{Fraction of the negative mobility area in the analysed parameter space, $a \in [0,20]$, $\omega \in [0,20]$ is presented in panel (a) for $D = 0$ and different values of the bias $f$ and (b) for $f = 0.5$ and various temperatures $D$.}
	\label{fig4}
\end{figure}
\subsection{Optimal mass for the presence of negative mobility}
Since the impact of the mass $m$ seems to be non-trivial, an interesting task is to find the value of $m$ for which the presence of negative mobility is the most common. Our numerical scans of the parameter space allowed us to determine this value. The result is shown in Fig. \ref{fig4}. In panel (a) we depict fraction of the negative mobility area in the analysed parameter space for the deterministic system $D = 0$ and different values of the bias $f$. A perhaps surprising finding is that for the small values of the static load $f = 0.25$ the optimal mass for the presence of negative mobility is $m \approx 0.13$. It is significantly less than the magnitude of dimensionless friction coefficient, which for the employed scaling is equal to unity $\gamma = 1$. This fact indicates that the friction plays prevalent role for the emergence of negative mobility.  Moreover, even for the very strongly damped Brownian particle $m \ll \gamma = 1$ the area of negative mobility is non-zero and relatively large. In both limiting regimes of the overdamped $m \to 0$ and the underdamped $m \to \infty$ motion there are no regions of  negative mobility. Ipso facto we confirmed numerically the no-go theorem formulated in Ref. \cite{speer2007}. This conclusion is valid also for larger values of the bias $f$, however, then the optimal mass $m$ for the presence of negative mobility becomes shifted towards higher, e.g. $m \approx 0.47$ for $f = 0.8$. Moreover, as the static load $f$ is increased the overall occurrence of negative mobility in the analysed parameter space is decreased and the presented curves come to be more flattened. In panel (b) we show the same characteristic but for the fixed bias $f = 0.5$ and different values of thermal noise intensity $D$. As temperature grows the regions of negative mobility in the parameter space tend to contract which is illustrated in the figure. Apart from this fact for stronger thermal noise, c.f. the case $D = 10^{-3}$, the curve becomes noticeably more bimodal. This observation is most likely due to parameter regimes for which negative mobility is induced by thermal fluctuations.

\begin{figure}[t]
	\centering
	\includegraphics[width=1.0\linewidth]{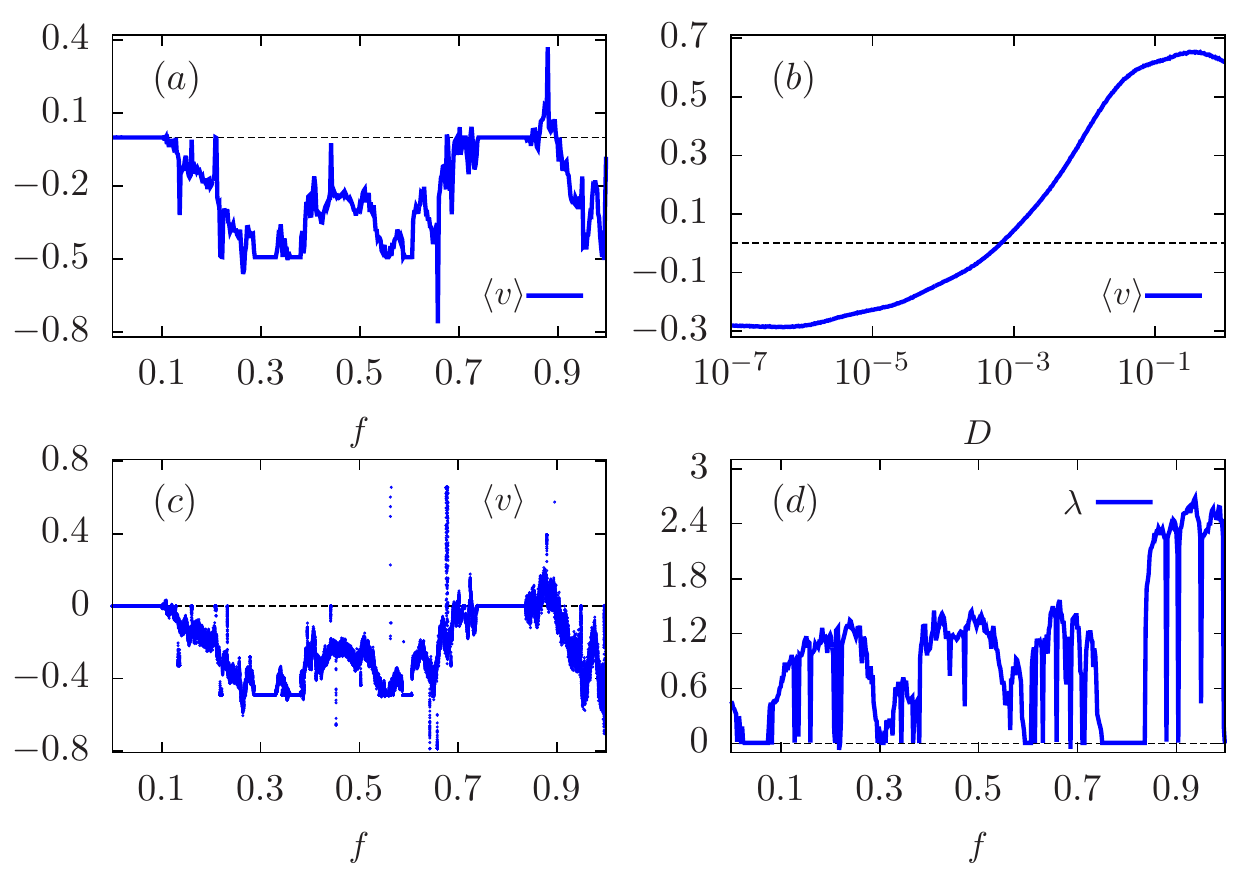}
	\caption{The negative mobility of the strongly damped Brownian particle $m \ll 1$ induced by the deterministic chaotic dynamics. Panel (a) the directed velocity $\langle v \rangle$, (c) bifurcation diagram of the directed velocity $\langle v \rangle$, (d) the maximal Lyapunov exponent $\lambda$ as the function of the external static bias $f$ with $D = 0$. Panel (b) the directed velocity $\langle v \rangle$ versus thermal noise intensity $D$ for $f = 0.66$. Other parameters are $m = 0.0555$, $a = 8.55$, $\omega = 12.38$.}
	\label{fig5}
\end{figure}
\subsection{The mechanisms of negative mobility}
To gain further insight into the nature of negative mobility in this system, as the next step we identify, exemplify and analyse the mechanisms standing behind emergence of this phenomenon. In Fig. \ref{fig5} we present the regime of parameters for which negative mobility occurs on grounds which are rooted solely in the complex, strongly damped and deterministic chaotic dynamics. Panel (a) depicts the directed velocity $\langle v \rangle$ versus the external static bias $f$. It is a nonlinear function without any obvious relation to the magnitude of the force $f$. Clearly, there are two windows of the latter parameter for which negative mobility $\langle v \rangle < 0$ is observed. The first starts at $f \approx 0.1$ and ends at $f \approx 0.7$. The second is present for the bias larger than approximately $f \approx 0.9$. In panel (b) we present the directed velocity $\langle v \rangle$ of the Brownian particle as a function of thermal noise intensity $D$ for the fixed bias $f = 0.66$ which corresponds to the sharp minimum of $\langle v \rangle$ depicted in the panel (a). In the limiting case of the very low temperatures $D \to 0$ the measured directed velocity is less than zero indicating that  negative mobility has its origin in the complex deterministic dynamics. For increasing temperature the directed velocity grows as well up to the critical thermal noise intensity $D \approx 10^{-3}$, for which the Brownian particle response reverses its sign $\langle v \rangle > 0$. In the high temperature limit $D \to \infty$ (not depicted) all forces in the right hand side of Eq. (\ref{dimlessmodel}) become negligible in comparison to thermal noise and thus the directed velocity vanishes completely $\langle v \rangle = 0$. Panel (c) of the same figure presents the bifurcation diagram of the latter quantity $\langle v \rangle$ illustrated as the function of the external bias $f$ for the deterministic system with $D = 0$. Each blue dot represents an attractor for the asymptotic long time directed velocity $\langle v \rangle$. For almost all values in the considered range of the bias $f$ there is the continuum of the directed velocity solutions. This fact suggests that the system is predominantly chaotic in this interval. We confirm this hypothesis in panel (d) where we depict the maximal Lyapunov exponent $\lambda$ for the deterministic system described by Eq. (\ref{dimlessmodel}) with $D = 0$ versus the biasing force $f$. Accordingly, this quantity is positive in almost entire considered interval of the parameter $f$. In particular, it is so for the values of $f$ corresponding to negative mobility. Therefore, we conclude that in the presented parameter regime this phenomenon is induced solely by the chaotic deterministic dynamics of the system given by Eq. (\ref{dimlessmodel}). Such a mechanism has been already reported in literature \cite{machura2007,speer2007}, however, here we prove that it may operate also for the strongly damped Brownian particle $m \ll \gamma = 1$.

\begin{figure}[t]
	\centering
	\includegraphics[width=1.0\linewidth]{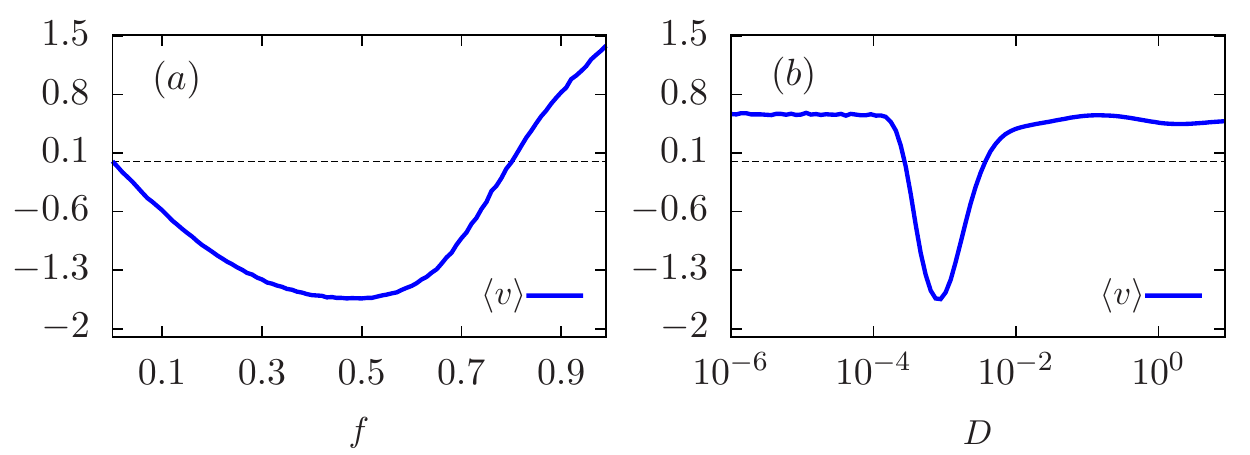}
	\caption{The negative mobility of the strongly damped Brownian particle $m \ll 1$ induced by thermal equilibrium fluctuations. The directed velocity $\langle v \rangle$ is presented versus external static bias $f$ in panel (a) and versus thermal noise intensity $D$ in panel (b). Parameters are the same as in Fig. \ref{fig5}, except now $m = 0.1047$, $f = 0.5$ and $D = 0.0009$.}
	\label{fig6}
\end{figure}
The second mechanism of the emergence of negative mobility is exemplified in Fig. \ref{fig6}. In panel (a) we present the directed velocity $\langle v \rangle$ of the strongly damped Brownian particle $m \ll \gamma = 1$ versus the external static bias $f$ for thermal fluctuations intensity $D = 0.0009$. In this case a very small amount of noise yields negative mobility in the linear response regime, i.e. for small forces $f$. For larger values of the bias $f$ the directed velocity is positive $\langle v \rangle > 0$ and the particle moves in the direction pointed by the static force $f$. Interestingly, there is an optimal value of $f \approx 0.5 $ for which negative mobility is most pronounced. To gain further insight into the origin of the discussed anomaly in the presented regime, in the neighbouring panel (b) we study the directed velocity $\langle v \rangle$ as a function of thermal noise intensity $D \propto T$. Contrary to the previously presented case, here at  low temperature $D \to 0$ the Brownian particle velocity is positive. The above described negative mobility manifests itself only in finite interval of temperature $D\in (2.8\times 10^{-4}, 3.2\times 10^{-3})$.  Further increase of thermal noise intensity leads to disappearance of this phenomenon. Although a solely noise induced negative mobility can occur only under impact of thermal fluctuations, the underlying relevant mechanism is strongly influenced by the deterministic dynamics as it was already shown in Ref. \cite{machura2007}.

\begin{figure}[t]
	\centering
	\includegraphics[width=1.0\linewidth]{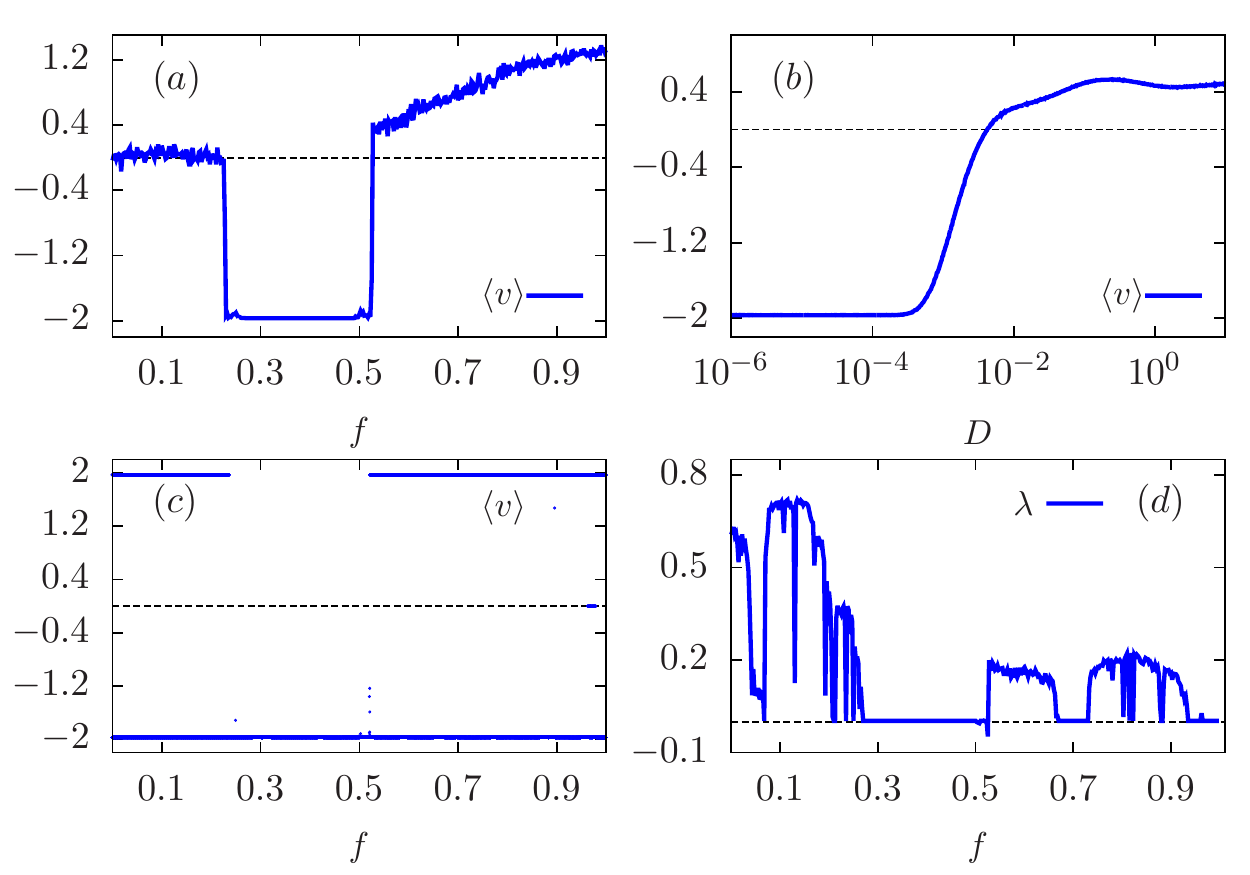}
	\caption{The negative mobility of the strongly damped Brownian particle $m \ll 1$ induced by the deterministic non-chaotic dynamics. Panel (a) the directed velocity $\langle v \rangle$, (c) bifurcation diagram of the directed velocity $\langle v \rangle$, (d) the maximal Lyapunov exponent $\lambda$ as the function of the external static bias $f$ with $D = 0$. Panel (b) the directed velocity $\langle v \rangle$ versus thermal noise intensity $D$ for $f = 0.5$. Other parameters are the same as in Fig. \ref{fig5} and $m = 0.1$.}
	\label{fig7}
\end{figure}
Finally, in Fig. \ref{fig7} we present a case of negative mobility induced by the deterministic {\it non-chaotic} dynamics of the system. Panel (a) illustrates the directed velocity $\langle v \rangle$ as a function of the external bias $f$ for the deterministic case $D = 0$ and in the strong  damping regime  $m \ll 1$. For very small values of the force $f$ the directed velocity oscillates around zero. When the bias is of moderate magnitude there is a window for which the particle response is opposite to the applied constant perturbation, so we detect there negative mobility. Further increase of the external force rapidly reverses the particle current and causes its monotonic growth. In panel (b) we study the impact of thermal fluctuations on $\langle v \rangle$ in the parameter regime with $f = 0.5$ corresponding to the minimal plateau depicted in the plot (a). Indeed, for the deterministic limit of the dynamics $D \to 0$ the directed velocity $\langle v \rangle$ is negative suggesting that the observed phenomenon of negative mobility has the deterministic origin. An increase of thermal noise intensity ceases this effect. For this parameter regime there is a surprisingly simple structure of attractors for the directed velocity which is visualized in the panel (c) in the deterministic case with $D = 0$. In the considered interval of the external force $f$ there are two asymptotically stable solutions corresponding to $\langle v \rangle = \pm 2$. Notably, in the bias window where negative mobility is observed only the attractor $\langle v \rangle = -2$ survives. This unexpected simplicity of solutions suggests that in the considered parameter regime the deterministic dynamics is non-chaotic, nonetheless still exhibits negative mobility. Our finding is confirmed in panel (d) where we depict the maximal Lyapunov exponent $\lambda$ versus the bias $f$ for the system with $D = 0$. An interesting observation is that the dynamics is generally chaotic ($\lambda > 0$) when two attractors $\langle v \rangle = \pm 2$ coexist and it is non chaotic with $\lambda = 0$ in the window where negative mobility emerges. To the best of the authors knowledge such a mechanism has never been reported. It may potentially open a possibility of observation of the negative mobility phenomenon for the discontinuous or non-ergodic one-dimensional nonequilibrium overdamped dynamics, corresponding to the formal substitution $m = 0$ in Eq. (\ref{dimlessmodel}) \cite{speer2007}.

\begin{figure}[t]
	\centering
	\includegraphics[width=1\linewidth]{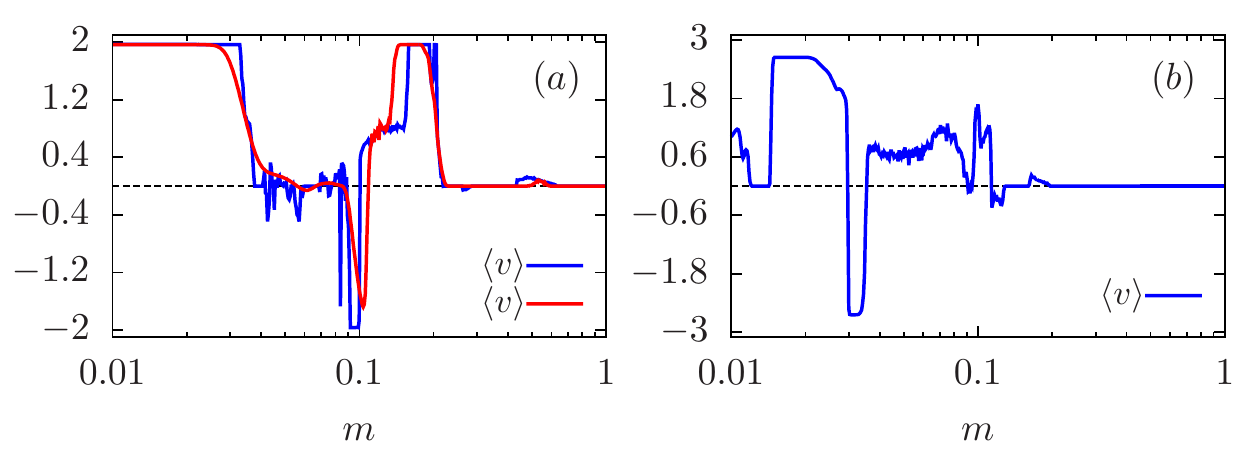}
	\caption{The directed velocity $\langle v \rangle$ of the driven Brownian particle versus its inertia $m$. Panel (a): $a = 8.55$, $\omega = 12.38$ and $f = 0.5$. The blue curve is for the deterministic case $D=0$ and the red curve is for the  noisy system with $D=0.0009$. Panel (b): $a=9.845$, $\omega = 16.64$, $f=0.25$ and $D=10^{-5}$.}
	\label{fig8}
\end{figure}	
\subsection{Impact of inertia}
To conclude this section we present in Fig. \ref{fig8} the representative dependence of the directed velocity $\langle v \rangle$ on the inertia $m$ for the deterministic and noisy system.  In panel (a) the amplitude $a$ and frequency $\omega$ of the periodic driving are the same as in the previous figures 5-7. Here, the most pronounced negative mobility is observed for the mass $m\approx 0.1$. For specifically tailored parameter sets this phenomenon could be detected for even smaller mass. We exemplify this situation in panel (b) where it is observed for $m\approx 0.03$  which is indeed the regime of very strong damping. 

We observe that the system response is very sensitive to even smallest changes of the particle mass $m$. Moreover, there are multiple reversals of the sign of the directed velocity which are characteristic for a massive setup driven by the external harmonic force \cite{bartussek1994,jung1996}. This finding can be utilized to particle sorting \cite{eichhorn2010}. For instance, one can see from the above figure that particles with different masses can easily be guided into opposite direction by a suitable choice of the system parameters. In addition, we want to point out that the occurrence of the three presented mechanisms of negative mobility are controlled by the magnitude of the particle inertia. Depending on its value the deterministic chaotic, thermal noise induced and the deterministic non-chaotic anomalous transport can be observed. It is worth to explicitly note that a tiny change from $m = 0.1$ to $m = 0.1047$ transforms the nature of negative mobility effect from deterministic non-chaotic to thermal noise induced, c.f. Fig. \ref{fig6} and Fig. \ref{fig7}. Therefore, the outlined mechanisms of the occurrence of negative mobility are also very sensitive to alteration of the particle inertia.

\section{Conclusion}
In this work we investigated the impact of inertia on  transport properties of a  Brownian particle moving in a periodic symmetric structure,  which in addition is exposed to a harmonic ac driving as well as a constant bias. The  parameter space is 5-dimensional and its complete numerical  exploration is far beyond the scope of this work. 
We first analysed the general behaviour of the  directed velocity $\langle v \rangle$ as a function of the amplitude $a$ and the angular frequency $\omega$ of the driving, for selected values of the remaining system parameters: the particle inertia $m$, the external bias $f$ and thermal noise intensity $D$. These results reveal especially that the negative mobility phenomenon emerges also for the strongly damped motion of a Brownian particle when the dissipation dominates over inertia.   Our scans of the parameter space allowed us to determine the optimal mass $m \approx 0.35$ for the  presence of the negative mobility phenomenon. By observing that the fraction of the negative mobility area in the analysed parameter space disappears for both the limiting cases of the overdamped $m \to 0$ and underdamped $m \to \infty$ motion, we confirmed with precise numerics the no-go theorem formulated in Ref.  \cite{speer2007}. We gained further insights into the origin of negative mobility in this system by revealing three classes of mechanisms responsible for this anomalous transport process. It can (i) be caused by the complex deterministic chaotic dynamics of the system or (ii) induced by the thermal noise, or (iii) associated with the deterministic, yet non-chaotic system evolution. In particular, according to the best of authors knowledge, the latter origin has never been reported before. It may provide guidance to the possibility of observation of the negative mobility phenomenon for the discontinuous or non-ergodic one-dimensional nonequilibrium overdamped dynamics when the particle inertia is fixed to zero $m = 0$. This case is of fundamental importance from the point of view of biological systems, all of which in situ operate in strongly fluctuating environments. Finally, we depict the illustrative impact of the particle inertia on its transport properties to unravel its spectacular sensitiveness to variation of this parameter. We detect the phenomenon of multiple velocity reversals which may constitute a cornerstone of particle sorting. Moreover, a small change in the particle inertia can radically alter the mechanism of negative mobility.

Our results can readily be experimentally tested with a single Josephson junction device or cold atoms moving in an dissipative optical lattice. Suitable parameter values, for which the above effects are predicted to occur, are accessible experimentally.

\section{Acknowledgement}
The work was supported by the NCN grant 2015/19/B/ST2/02856 (J. S. \& J. {\L}.).

\end{document}